\documentclass[a4paper,10pt,twocolumn]{article}
\usepackage{graphicx}
\usepackage{xcolor}
\usepackage{url}
\usepackage[colorlinks=true,linkcolor=blue,citecolor=blue]{hyperref}
\usepackage[expproduct=cdot]{siunitx} 

\usepackage{amsmath}
\usepackage[charter]{mathdesign}

\usepackage{ctable}\newcommand{\thickmidrule}{\midrule[\heavyrulewidth]} 
\usepackage{tabularx} 

\usepackage{cite} 

\definecolor{royalblue4}{HTML}{27408B}
\definecolor{red4}{HTML}{8B0000}
\definecolor{green4}{HTML}{008b00} 

\usepackage{dblfloatfix} 
\usepackage{fixltx2e} 

\usepackage[font=footnotesize,labelfont=bf,labelsep=endash,margin=0pt]{caption} 
\usepackage[title,header]{appendix}

\newlength{\myleftmargin} 
\newlength{\mytopmargin} 
\newlength{\myrightmargin} 
\newlength{\mybottommargin} 
\usepackage[runin]{abstract}

\newcommand{\keywords}[1]{\vspace{2mm}\noindent\textbf{Key words:} #1} 
\newcommand{\pagewidetitle}[3] 
{%
    \twocolumn%
        [%
            \vskip-5mm%
            \begin{@twocolumnfalse}%
                #1%
                #2%
                \vspace{5mm}%
            \end{@twocolumnfalse}%
        ]%
        #3%
}

\newlength{\figurewidth}

\newcommand{\placeobj}[4][l]{%
    \makebox[0pt][#1]{\hspace{#2}\raisebox{-#3}[0pt][0pt]{#4}}%
}

\newcommand{\ie}{{\it i.e.}}

\newcommand{\eg}{{\it e.g.}}
\newcommand{\etal}{{\it et\ al.}}
\newcommand{\etc}{{\it etc}}

\newcommand{\vs}{\textit{vs.}}

\renewcommand{\d}{\mathrm{d}}
\newcommand{\p}{\partial}

\newcommand{\Order}{\mathrm{O}}

\newcommand{\e}{\mathrm{e}}

\newcommand{\da}{\ensuremath{\text{day}}} 
\newcommand{\days}{\ensuremath{\text{days}}} 

\newcommand{\um}{\ensuremath{\micro\metre}}
\newcommand{\pM}{\ensuremath{\text{pM}}} 

\usepackage{relsize} 
\newcommand{\bmu}{\textsc{bmu}} 

\newcommand{\ob}{\textsc{ob}}

\newcommand{\msc}{\textsc{msc}} 
\newcommand{\obu}{\text{\textsc{ob}$_\text{u}$}} 

\newcommand{\obp}{\text{\textsc{ob}$_\text{p}$}}
\newcommand{\obpsat}{\ensuremath{\ob_\text{p}^\text{sat}}}
\newcommand{\oba}{\text{\textsc{ob}$_\text{a}$}}

\newcommand{\ocp}{\text{\textsc{oc}$_\text{p}$}}

\newcommand{\oca}{\text{\textsc{oc}$_\text{a}$}}

\newcommand{\tgfb}{\textsc{tgf\textsmaller{$\betaup$}}}
\newcommand{\Tgfb}{\textsc{Tgf$\betaup$}}

\newcommand{\igf}{\text{\textsc{igf}}}
\newcommand{\pge}{\text{\textsc{pge}\textsmaller{2}}}
\newcommand{\wnt}{\text{\textsc{w}\textsmaller{nt}}}
\newcommand{\Wnt}{\textsc{W}\textsmaller{nt}}

\def\dkk#1{\text{\textsc{d}\textsmaller{kk#1}}}
\def\sfrp#1{\textsc{\textsmaller{s}frp\textsmaller{#1}}}

\newcommand{\rank}{\textsc{rank}}
\newcommand{\rankl}{\textsc{rankl}}

\newcommand{\opg}{\textsc{opg}}

\def\pgde2{\text{\textsc{pgde}$_2$}}
\newcommand{\pth}{\textsc{pth}}
\newcommand{\prodextpth}{\ensuremath{P^\text{ext}_\pth}}
\newcommand{\pthrp}{\text{\textsc{pth}\textsmaller{r}\textsc{p}}}
\newcommand{\pthrpcleaved}{\text{\textsc{pth}\textsmaller{r}\textsc{p}\textsmaller{[1--23]}}}
\def\lrp#1{\text{\textsc{lrp}\textsmaller{#1}}}
\newcommand{\piact}{\ensuremath{\pi^\text{act}}} 
\newcommand{\pirep}{\ensuremath{\pi^\text{rep}}} 
\newcommand{\bv}{\textsc{bv}}

\newcommand{\kform}{\text{$k_\text{form}$}} 
\newcommand{\kres}{\ensuremath{k_\text{res}}}

\newcommand{\psa}{\textsc{psa}} 
\newcommand{\pca}{\text{\textsc{pc}\textsmaller{a}}} 
\newcommand{\pcamax}{\text{\textsc{pc}\textsmaller{a}$^\text{max}$}} 

\newcommand{\dobu}{\ensuremath{\mathcal{D}_\obu}}
\newcommand{\dobp}{\ensuremath{\mathcal{D}_\obp}}
\newcommand{\aoba}{\ensuremath{\mathcal{A}_\oba}}
\newcommand{\pobp}{\ensuremath{\mathcal{P}_\obp}}
\newcommand{\docp}{\ensuremath{\mathcal{D}_\ocp}}
\newcommand{\aoca}{\ensuremath{\mathcal{A}_\oca}}

\newcommand{\dobubar}{\ensuremath{\overline{\mathcal{D}}_\obu}}
\newcommand{\dobpbar}{\ensuremath{\overline{\mathcal{D}}_\obp}}

\newcommand{\pobpbar}{\ensuremath{\overline{\mathcal{P}}_\obp}}

\newcommand{\obpbar}{\ensuremath{\overline\obp}}
\newcommand{\obabar}{\ensuremath{\overline\oba}}
\newcommand{\ocabar}{\ensuremath{\overline\oca}}
\newcommand{\tgfbbar}{\ensuremath{\overline\tgfb}}
\newcommand{\ranklbar}{\ensuremath{\overline\rankl}}
\newcommand{\sigmabar}{\ensuremath{\overline\sigma}}

\begin{document}
\title{\bf Modelling the anabolic response of bone using a cell population model}
\author{Pascal R. Buenzli$^\text{1}$, Peter Pivonka, Bruce S. Gardiner, David W. Smith}
\date{\small \vspace{-2mm}Faculty of Engineering, Computing \& Mathematics,
    \\The University of Western Australia, WA 6009, Australia\\\vskip 1mm \normalsize \today\vspace*{-5mm}}

\pagewidetitle{
\maketitle}{
\begin{abstract}
    To maintain bone mass during bone remodelling, coupling is required between bone resorption and bone formation. This coordination is achieved by a network of autocrine and paracrine signalling molecules between cells of the osteoclast lineage and cells of the osteoblastic lineage. Mathematical modelling of signalling between cells of both lineages can assist in the interpretation of experimental data, clarify signalling interactions and help develop a deeper understanding of complex bone diseases. Several mathematical models of bone cell interactions have been developed, some including \rank--\rankl--\opg\ signalling between cells and systemic parathyroid hormone \pth. However, to our knowledge these models do not currently include key aspects of some more recent biological evidence for anabolic responses. In this paper, we further develop a mathematical model of bone cell interactions by Pivonka \etal\ (2008)~\cite{pivonka-etal-1} to include the proliferation of precursor osteoblasts into the model. This inclusion is important to be able to account for \wnt\ signalling, believed to play an important role in anabolic responses of bone. We show that an increased rate of differentiation to precursor cells or an increased rate of proliferation of precursor osteoblasts themselves both result in increased bone mass. However, modelling these different processes separately enables the new model to represent recent experimental discoveries such as the role of \wnt\ signalling in bone biology and the recruitment of osteoblast progenitor cells by transforming growth factor~$\betaup$. Finally, we illustrate the power of the new model's capabilities by applying the model to prostate cancer metastasis to bone. In the bone microenvironment, prostate cancer cells are believed to release some of the same signalling molecules used to coordinate bone remodelling (\ie\ \wnt\ and \pthrp), enabling the cancer cells to disrupt normal signalling and coordination between bone cells. This disruption can lead to either bone gain or bone loss. We demonstrate that the new computational model developed here is capable of capturing some key observations made on the evolution of the bone mass due to metastasis of prostate cancer to the bone microenvironment.

\end{abstract}
}{
\protect\footnotetext[1]{Corresponding author. Email address: \texttt{pascal.buenzli@uwa.edu.au}}
}

\section{Introduction}
Bone is a dynamic living tissue which continuously undergoes remodelling to ensure mineral homeostasis and to repair micro damage~\cite{parfitt1,martin-burr-sharkey}. The two main bone cell types executing bone remodelling are osteoclasts which resorb the mineralised bone matrix and osteoblasts which deposit osteoid (which subsequently becomes mineralised)~\cite{martin-burr-sharkey}. The third cell type involved in bone remodelling are osteocytes (\ie, terminally differentiated cells derived from mature osteoblasts that have been trapped in the mineralised bone matrix~\cite{bonewald-johnson}). The entire ensemble of bone cells contributing to bone remodelling is referred to as basic multicellular unit (\bmu)~\cite{parfitt3,parfitt-in-recker}.

Within the \bmu, pre-osteoblasts, which express \rankl\ have been hypothesised to control the differentiation of osteoclasts from hematopoietic progenitors~\cite{ma-martin-etal,martin,roodman,gori-etal}. The bone resorption phase is subsequently followed by bone formation, driven in part, by factors produced by the osteoclast that stimulate osteoblastogenesis~\cite{roodman}. This coupling between resorption and formation phase in \bmu s is required to maintain bone mass. Many bone pathologies, such as osteoporosis, Paget's disease and cancer metastasis to bone, are associated with the dysregulation of this coupling process leading to abnormal bone loss or bone gain. Mathematical modelling can be employed to interpret experimental data, clarify signalling interactions, investigate therapeutic interventions, and to generally better understand bone remodelling from a systems perspective~\cite{pivonka-komarova}.

Bone remodelling has been represented mathematically in a variety of ways including bone cell population models (ODEs)~\cite{komarova-etal1,lemaire-etal,pivonka-etal-1}, continuum models (PDEs)~\cite{ryser-etal1,buenzli-pivonka-smith,ji-etal} and discrete cell models~\cite{vanOers-etal,buenzli-jeon-etal}. The bone cell population model by Lemaire \etal~\cite{lemaire-etal} proposes an interesting approach based on fundamental chemical reaction principles such as material balance and mass action kinetics. This model incorporates some of the most important bone biology known at that time. Extensions to include further components of bone biology can be formulated using the same framework. We have used this framework to include new knowledge in bone biology in our bone cell population model~\cite{pivonka-etal-1} (such as the expression of \rankl\ and \opg\ by osteoblasts of various maturities)~\cite{pivonka-etal-1}, and to introduce a spatial variation in cells numbers to represent a single basic multicellular unit~\cite{buenzli-pivonka-smith}. We have also applied the model by Pivonka \etal~\cite{pivonka-etal-1} to examine possible therapeutic interventions to restore bone mass following dysregulation of the \rank--\rankl--\opg\ signalling system~\cite{pivonka-etal-2}, coupled this model to a pharmacokinetic model of denosumab to explore the effect of different dosing regimes~\cite{scheiner-etal}, and studied osteolytic lesions in multiple myeloma~\cite{wang-etal-bone-mm}.

However, while the model by Pivonka~\etal~\cite{pivonka-etal-1} does some things well, it does not capture the anabolic effects of precursor osteoblast proliferation. Recent experimental evidence suggests that \wnt\ signalling is a critically important regulator of bone remodelling---\wnt\ signalling plays an important role in normal bone homeostasis under varying mechanical loading, and excessive \wnt\ signalling is responsible for some osteopetrotic (excess) bone states~\cite{henriksen-etal,jilka}. In addition, recent clinical evidence demonstrates that administration of intermittent \pth\ is an effective anabolic intervention~\cite{jilka,hodsman-etal}. The exact molecular mechanisms leading to anabolic responses under intermittent \pth\ administration are incompletely understood and probably multifactorial, involving differential regulations of osteoblast differentiation, proliferation and apoptosis~\cite{jilka}. While we do not model intermittent \pth\ administration in this paper, it is important to include these three cellular behaviours regulating the number of osteoblasts for future investigations. In this paper, we thus further develop the model by Pivonka et al~\cite{pivonka-etal-1} by introducing the proliferation of osteoblasts in a way such that the new model is consistent with the original model and can incorporate osteoblast proliferation through \wnt\ signalling or via other signalling systems. We then explore the effect of parameter changes in the new model on net bone balance, and see that the new model is capable of effectively representing osteopetrotic bone disease states arising from disruption of normal osteoblastic proliferation. 

Finally we illustrate the capabilities of the new model in a complex bone disease that arises when prostate cancer cells metastasise to the bone microenvironment. This disease is characterised by a variable phenotype that often involves net bone gain (coupled with focal bone loss)~\cite{clarke-fleisch,hall-etal-2006,roudier-etal}, and finally net bone loss. We show that the new model developed here can model bone gain and bone loss via secretion of signalling molecules such as \wnt, \psa\ and \pthrp\ by the prostate cancer cells.

\section{Background}
\begin{figure*}
    \centering\includegraphics[width=0.7\textwidth]{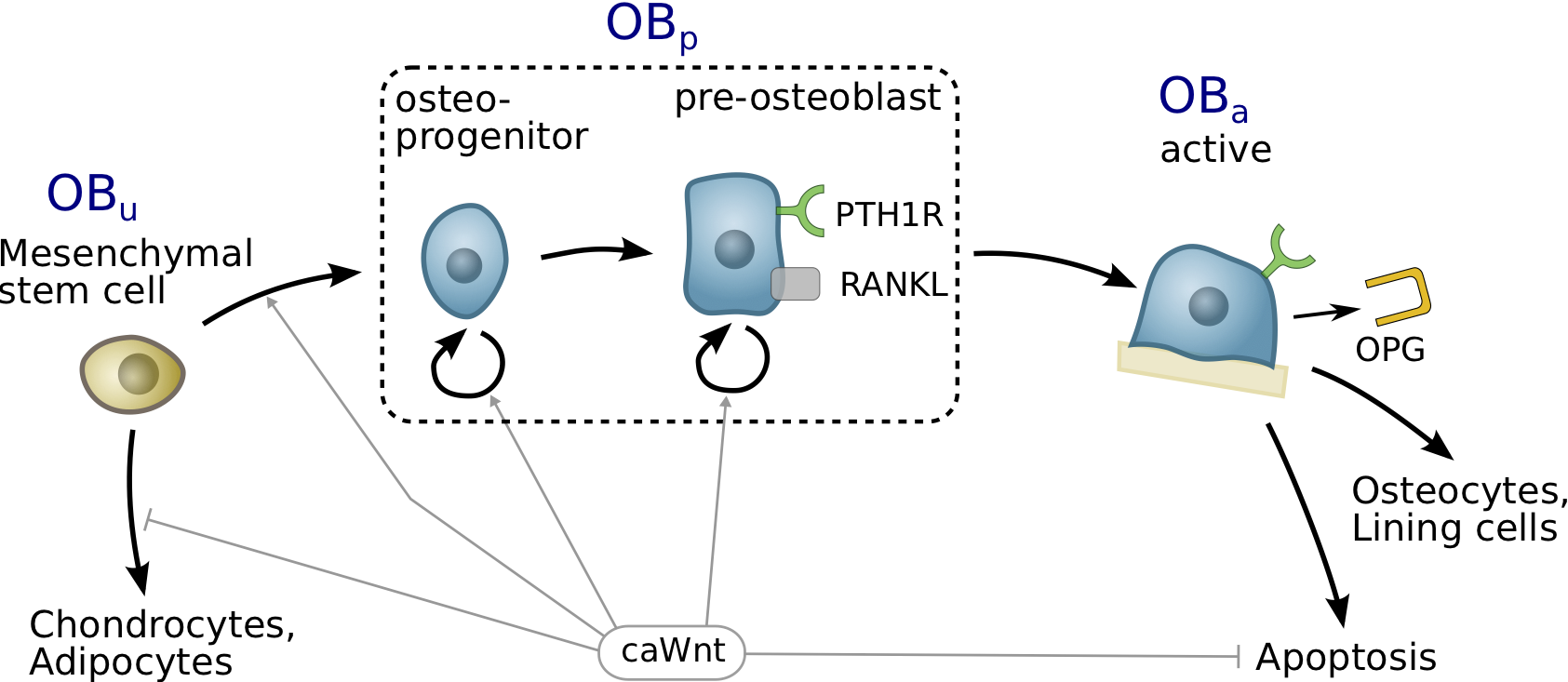}
    \caption{Osteoblast development and the \wnt\ signalling pathway. Canonical \wnt\ promotes the commitment of \msc s to the osteoblastic lineage, stimulates osteoblast proliferation and enhances osteoblast and osteocyte survival~\cite{khosla-etal}.}
    \label{fig:osteoblast-development}
\end{figure*}

A recent review by Khosla~\etal\ highlights the importance of osteoblast development in the regulation of bone remodelling and the potential for therapeutic interventions that target the osteoblastic lineage~\cite{khosla-etal}. Osteoblasts are mesenchymal cells derived from the mesoderm. Sequential expression of several molecules (such as \textsc{r\textsmaller{unx2}} and \textsc{o\textsmaller{sx}}), driven by signal transduction pathways, facilitates the differentiation of the progenitor cell into a proliferating pre-osteoblast, then into a bone matrix-producing osteoblast, and eventually into a mechanosensory osteocyte or a bone-lining cell (see Figure~\ref{fig:osteoblast-development}). As the cells of the osteoblastic lineage differentiate, they produce molecules essential for regulating \bmu\ operation, including support of osteoclastogenesis and angiogenesis in a \bmu. Active osteoblasts secrete osteoid, which later mineralises to bone, while osteocytes produce molecules that regulate \bmu\ function.

The most extensively studied cell-kinetic model of osteoblast development is that of mechanically induced bone formation in experimental orthodontics \cite{kimmel-jee,roberts-mozsary-klinger,mcculloch-melcher}. Based on nuclear size, Roberts \etal~\cite{roberts-mozsary-klinger} have characterised four precursor cell types to the functional osteoblast. This differentiation pathway has been confirmed (and refined) by marker expression \textit{in vitro} using functional assays~\cite{aubin-bookchapter,liu-etal}. Asymmetrically-dividing mesenchymal stem cells recruited to the \bmu\ give rise to a population of osteoblast progenitors that are proliferating extensively (undergoing symmetric division). These osteoblast progenitors differentiate into pre-osteoblasts that undergo limited proliferation. Finally, pre-osteoblasts differentiate in turn into non-proliferative active osteoblasts~\cite{aubin-bookchapter}.

Current bone biology literature identifies the central role played by the \wnt\ signalling pathway in regulating osteoblast development (Figure~\ref{fig:osteoblast-development}). \Wnt s are a family of over 20 secreted glycoproteins crucial for the development and homeostatic renewal of many tissues, including bone~\cite{fuerer-etal}. \Wnt s stimulate canonical or non-canonical signalling pathways by binding a receptor complex consisting of \textsc{ldl} receptor-related protein 5 (\lrp5) or \lrp6 and one of ten Frizzled (\textsc{f}\textsmaller{z}) molecules~\cite{uitterlinden-etal}. There are a range of soluble `decoy molecules' secreted that regulate \wnt\ signalling including sclerostin, \dkk{1,2,3} and \sfrp{1,2,3}. The canonical \wnt\  signalling pathway has been the most extensively studied \wnt\ signalling pathway in osteoblasts. \Wnt\ activation involves the stabilization of $\betaup$-catenin (via inhibition of the \textsc{gsk-\textsmaller{3$\betaup$}}, axin and \textsc{apc} complex), resulting in the translocation of $\betaup$-catenin \textsc{tcf/lef} to the cell nucleus and activation of various downstream transcription factors critical for directing cell lineage and subsequent cell proliferation~\cite{krishnan-etal}. \Wnt\ signalling has three major functions in osteoblastic lineage cells: (i) dictating osteoblast specification from osteo-/chondroprogenitors; (ii) stimulating osteoblast proliferation; and (iii) enhancing osteoblast and osteocyte survival (Figure~\ref{fig:osteoblast-development}).

This brief overview of osteoblast development and the importance of \wnt\ signalling highlights the complexity of potential bone cell interactions. A systems biology approach to bone remodelling can help understand these interconnections and their importance for functional bone remodelling~\cite{pivonka-komarova}. It is only recently that a few mathematical models of interacting bone cells have been developed to explore these fundamental aspects of the bone remodelling sequence.

In Lemaire \etal~\cite{lemaire-etal}, a bone cell population model for bone remodelling is proposed and applied to the study of bone diseases and therapeutic strategies. To restore bone mass following catabolic pathologies (such as due to estrogen deficiency, vitamin~D deficiency, and senescence), the generation of pre-osteoblasts by differentiation is shown by this model to be a powerful bone formative strategy. This occurs despite coupling of formation to resorption through cells of osteoblastic lineage expressing \rankl, which binds to the \rank\ receptor of osteoclasts thereby promoting osteoclast activation and bone resorption. The computational model of bone remodelling of Lemaire \etal\ has been refined by Pivonka \etal~\cite{pivonka-etal-1} who investigated the effect of \rankl\ and \opg\ expression profiles on cells of the osteoblastic lineage.

The aim of this paper is to investigate the effects of different developmental stages of osteoblasts on bone remodelling. For this purpose, we include a transient-amplifying (\ie\ proliferating) stage in the osteoblastic lineage in the bone cell population model of Pivonka \etal~\cite{pivonka-etal-1}. The motivation for this inclusion is twofold:
\begin{enumerate}
\item It is known that the density of \msc s in bone marrow is very low, and though \msc s are recruited to the \bmu\ site by \tgfb~\cite{tang-etal}, it is likely the models of Refs~\cite{lemaire-etal,pivonka-etal-1} rely on an unphysiological recruitment of a sufficient number of pre-osteoblasts for the stimulation of a sustained formative response. The fast increase in osteoblast population required in a \bmu\ remodelling event is believed to involve proliferative cells (undergoing symmetric cell division), \ie, so-called transient-amplifying osteoblast progenitors~\cite{manolagas-kousteni-jilka}.

\item Hormones and cytokines regulate stages of osteoblast development differently. The transient-amplifying stage of osteoblasts is known to be strongly dependent on various hormones, growth factors and other molecules, such as \wnt, insulin-like growth factor (\igf), prostaglandin \textsc{e}\textsmaller{2} (\pge)\ and estrogen~\cite{digregorio-etal,manolagas-kousteni-jilka,ogita-etal}. The anabolic effect of intermittent \pth\ is believed to operate through multifactorial regulation of osteoblast generation at several developmental stages~\cite{jilka}. Also, in some bone diseases, signalling by these hormones and growth factors is disrupted, leading to an abnormal population of osteoblasts and a subsequent imbalance of bone during remodelling. For these reasons, a more accurate account of the transient-amplifying stage of osteoblasts is essential for the realistic modelling of such diseases.
\end{enumerate}
However, we find that the inclusion of \obp\ proliferation proposed in this paper has to be treated with some care as it can lead both to an unstable dynamic system (not converging to a steady state with finite cell densities) and to potentially unphysiological system behaviour, for example an anabolic response to continuous \pth\ administration, which is experimentally known to be catabolic. These issues are addressed in detail in this paper.

\section{Description of the model}\label{sec:model}
For simplicity, osteoblast progenitors and pre-osteoblasts are pooled into a single proliferative cell type in our model, which we call \emph{pre-osteoblast} and denote by \obp. Three stages of osteoblast development and two stages of osteoclast development are included in the cell population model.

\paragraph{Osteoclasts.}
Pre-osteoclasts (\ocp s) represent circulating cells of hematopoietic origin. Pre-osteoclasts are assumed to mature into active osteoclasts (\oca s) upon activation of their \rank\ receptor by the ligand \rankl. Active osteoclasts are cells that resorb bone matrix at a rate $\kres$ assumed constant (in volume per unit time). In the model, \oca s are assumed to undergo apoptosis at a rate enhanced by the presence of \tgfb~ \cite{roodman,pivonka-etal-1,buenzli-pivonka-smith}. Thus, osteoclast development can be summed up schematically as:
\begin{align}
    \ocp\ \stackrel{\rankl\, +}{\longrightarrow}\ \oca\ \stackrel{\tgfb\, +}{\longrightarrow}\ \emptyset.
    \label{oc-development}
\end{align}

\paragraph{Osteoblasts.}
Uncommitted osteoblast progenitors (\obu s) represent a pool of \msc s. These \msc s are assumed to commit to the osteoblastic lineage by becoming pre-osteoblasts (\obp s) through activation of \tgfb\ signalling. In the model, \obp s represent transient-amplifying osteoblast progenitors and they are therefore assumed to proliferate. Their maturation into active osteoblasts (\oba s) is assumed to be downregulated by \tgfb. Active osteoblasts are cells that form bone matrix at a rate $\kform$ assumed constant (in volume per unit time). The fate of active osteoblasts is either (i) to be buried in osteoid and become osteocytes; (ii) to undergo apoptosis; or (iii) to become bone-lining cells covering the surface of newly-formed bone. In our model, the elimination of an \oba\ depletes the pool of matrix-synthesising cells and thereby includes all three possibilities. Thus, osteoblast development can be summed up schematically as:
\begin{align}
    \obu\ \stackrel{\tgfb\, +}{\longrightarrow}\  \obp\placeobj{-2.8ex}{-1.8ex}{\scalebox{1.25}{\rotatebox[origin=c]{180}{$\circlearrowright$}}}\ \stackrel{\tgfb\, -}{\longrightarrow}\ \oba\ \longrightarrow\ ...
    \label{ob-development}
\end{align}

\paragraph{Regulatory factors.} System-level coupling between the osteoclasts and osteoblasts occurs because the concentrations of the coupling signalling molecules \tgfb\ and of \rankl\ are themselves influenced by cellular actions. The growth factor \tgfb\ is assumed to be stored in the bone matrix and released into the microenvironment in active form by the resorbing \oca s~\cite{roodman,iqbal-sun-zaidi,tang-etal}. The ligand \rankl\ is assumed to be expressed on the surface of \obp s. However this expression can be blocked by binding to \opg, which in turn is assumed to be produced in soluble form by \oba s~\cite{gori-etal,thomas-etal}. The generation of \rankl\ and of \opg\ by osteoblasts is respectively upregulated and downregulated by the systemic hormone \pth.

\paragraph{Governing equations.}
The osteoclast and osteoblast development pathways~\eqref{oc-development}--\eqref{ob-development} are transcribed mathematically as so-called `rate equations' involving \ocp, \oca, \obu, \obp\ and \oba\ cell densities (number of cells per unit volume)~\cite{pivonka-etal-1,buenzli-pivonka-smith}:
\begin{align}
    \tfrac{\p}{\p t} \oca &= \docp \ocp - \aoca \oca, \label{oca}
    \\\tfrac{\p}{\p t} \obp &= \dobu \obu - \dobp \obp + \pobp \obp, \label{obp}
    \\\tfrac{\p}{\p t} \oba &= \dobp \obp - \aoba \oba, \label{oba}
\end{align}
where
\begin{align}
    &\docp(t) = D_\ocp \piact\big(\rankl(t)/k^\rankl_\ocp\big), \label{docp}
    \\&\aoca(t) = A_\oca \piact\big(\tgfb(t)/k^\tgfb_\oca\big), \label{aoca}
    \\&\dobu(t) = D_\obu \piact\big(\tgfb(t)/k^\tgfb_\obu\big),  \label{dobu}
    \\&\dobp(t) = D_\obp \pirep\big(\tgfb(t)/k^\tgfb_\obp\big). \label{dobp}
\end{align}
In Eqs~\eqref{oca}--\eqref{oba}, source and sink terms are specified according to transformation rates between cell types with first order reaction rates to account for the effect of population sizes. $\docp(t)$ is the differentiation rate of \ocp s into \oca s activated by \rankl, $\aoca(t)$ is the apoptosis rate of \oca s activated by \tgfb, $\dobu(t)$ is the differentiation rate of \obu s into \oba s activated by \tgfb, and $\dobp(t)$ is the differentiation rate of \obp s into \oba s repressed by \tgfb. The elimination rate of active osteoblasts, $\aoba$, is assumed unregulated and constant: $\aoba(t)\equiv A_\oba$. Activation and repression of these rates by  \rankl\ or \tgfb\ is expressed in Eqs.~\eqref{docp}--\eqref{dobp} in terms of the dimensionless functions
\begin{align}
    \piact(\xi) = \frac{\xi}{1+\xi}, \qquad \pirep(\xi) = 1-\piact(\xi) = \frac{1}{1+\xi}.
    \label{piact-pirep}
\end{align}
These functional forms of $\piact$ and $\pirep$ are based on the following assumptions.
Ligands such as $\rankl$ and $\tgfb$ modulate cell behaviours by binding to specific receptors on the cells and triggering intracellular signalling pathways. Following Refs~\cite{lemaire-etal,pivonka-etal-1,buenzli-pivonka-smith}, we assume that the signal received by a cell corresponds to the fraction of occupied receptors on the cell. This fraction is equal to $\piact(L/k)$, where $L$ is the extracellular ligand concentration and $k$ a binding parameter (dissociation binding constant)~\cite{lauffenburger-linderman}. We do not model intracellular pathways explicitly but relate a cell's response to its input signal by assuming a phenomenological relationship. Here, we assume that a cell responds in proportion to receptor occupancy, \ie\ either in proportion to $\piact$ (for activation) or to $\pirep$ (for repression).\footnote{Such a relationship has been shown to hold experimentally for example in the context of human fibroblasts stimulated by epidermal growth factor (EGF): the mitogenic response of these fibroblasts is linearly dependent on the fraction of occupied EGF receptors~\cite[Fig.\ 6-7, p.249]{lauffenburger-linderman}.} Note that since receptor occupancy is a nonlinear function of the free ligand concentration, the overall relationship between concentration of extracellular ligand $L$ and cell response in Eqs~\eqref{docp}--\eqref{dobp} is nonlinear.

The rate equations governing the concentrations of \tgfb, \rank, \rankl, \opg\ and \pth\ are solved under the approximation that receptor--ligand binding reactions occur on a fast timescale compared to cell responses. These equations are presented in Appendix~\ref{appx:model}.

The proliferation term $\pobp \obp$ in Eq.~\eqref{obp} has been added to the original system of equations of Ref.~\cite{pivonka-etal-1} to account for the transient-amplifying stage of osteoblasts. This term involves the proliferation rate $\pobp(t)$, which is related to the average cell cycle period of pre-osteoblasts, $\tau_\obp^\text{mitosis}$, by $\pobp(t) = \ln(2)/\tau_\obp^\text{mitosis}(t)$. The proliferation rate $\pobp(t)$ is controlled by a feedback mechanism and is therefore time dependent (see \emph{Regulation of \obp\ proliferation} below).

Finally, the matrix-resorptive activity of \oca s and matrix-synthesising activity of \oba s influence the overall amount of bone according to:
\begin{align}\label{bv}
    \tfrac{\p}{\p t} \bv = - \kres \oca + \kform \oba,
\end{align}
where $\bv$ stands for the volume fraction of bone matrix in a representative volume element at the tissue scale.\footnote{The volume fraction of bone matrix is also equal to $1-\Phi$ where $\Phi$ is the `bone porosity', \ie, the volume fraction of soft tissues (marrow, cells, stroma) (compare with Ref.~\cite[Eq.~(3.7)]{martin-burr-sharkey}).} The quantity $\kres \oca$ represents the resorption rate (bone volume fraction resorbed per unit time) and the quantity $\kform \oba$ represents the formation rate (bone volume fraction formed per unit time). All the parameter values of the model are listed in Appendix~\ref{appx:parameters} (Table~\ref{table:parameters}).

The system of ODEs~\eqref{oca}--\eqref{oba} together with Eqs.~\eqref{tgfb}--\eqref{pth} form a closed system that can be solved for the time evolution of the three state variables \obp, \oba\ and \oca\ from an initial condition. Eq.~\eqref{bv} can then be integrated to provide the time evolution of the bone volume fraction. Clearly, $\bv(t)$ is not a function of the current state only as it depends on the integrated history of $\oca(t)$ and $\oba(t)$. However, the bone volume fraction change rate $\tfrac{\p}{\p t}\bv$ is a function of the current state and will be a major model output followed in this paper.

\paragraph{Regulation of \obp\ proliferation.}
In a single \bmu, thousands of active osteoblasts refill the cavity created by the osteoclasts~\cite{martin-burr-sharkey,parfitt3}. Their continual recruitment from pre-osteoblasts occurs at a rate that varies with the rate of resorption. For a \bmu\ that advances in bone at $40\ \um/\da$, an estimated rate of 120~active osteoblasts per day is necessary to ensure that the whole perimeter of the \bmu\ cavity is covered by the bone refilling cells~\cite{martin-burr-sharkey}. This required recruitment rate of active osteoblasts is achieved by a combination of differentiation from mesenchymal stem cells near the tip of the blood vessel, and proliferation of pre-osteoblasts between the blood vessel and cavity walls~\cite{jaworski-hooper,roberts-mozsary-klinger,aubin-bookchapter}. Active osteoblasts in \bmu s usually form a single layer of cells~\cite{marotti-etal-1975}, and so are limited in number by the available bone surface area. It is likely that feedback control mechanisms regulate pre-osteoblast proliferation to limit the generation of active osteoblasts. We model this control of cell population by limiting the proliferation rate of pre-osteoblasts with the density of pre-osteoblasts, \ie:
\begin{align}
    \pobp(t) = \begin{cases} P_\obp(t) \left(1 - \frac{\obp(t)}{\obpsat}\right),\quad &\text{if}\ \obp(t) < \obpsat,
        \\0, \quad&\text{if}\ \obp(t) \geq \obpsat.\end{cases} \label{pobp}
\end{align}
In Eq.~\eqref{pobp}, $\obpsat$ is a critical density above which proliferation is entirely suppressed. The control of the proliferation rate by the density of \obp s may represent `contact inhibition' or `pressure inhibition' of proliferation and/or nutrient or space restrictions in the \bmu\ cavity. It may represent a feedback control from newly-formed active osteoblasts near the reversal zone.  Indeed, pre-osteoblasts represent the last stage of osteoblast development before maturation into active osteoblasts. The density of pre-osteoblasts at a given time used in Eq.~\eqref{pobp} is thus approximately proportional to the density of newly-formed active osteoblasts.

In Eq.~\eqref{pobp}, the remaining factor $P_\obp(t)$ stands for additional negative and\slash or positive regulations of the proliferation of \obp s as due to, \eg, \wnt\ signalling, \pth\ administration, and other hormones and growth factors that may play a role in mechanosensing~\cite{bonewald-johnson,scheiner-etal-mechanostat} or in the development of osteoporosis, viz.
\begin{align}
    P_\obp(t) \equiv P_\obp(\wnt(t), \pth(t), \text{estrogen}(t), \igf(t), \tgfb(t), ...).\label{pobp-further-regulations}
\end{align}
Including all these regulations is beyond the scope of the present work, but provides a clear direction for future research. Here we will assume that in normal bone homeostasis, systemic levels of these signalling molecules lead to a specific value of $P_\obp$ and we will first investigate how $P_\obp$ as a parameter affects the remodelling behaviour of the system. In Section~\ref{sec:pca}, the model is applied to a complex disease, and in this case, $P_\obp$ is made dependent on \wnt\ produced by metastatic prostate cancer cells.

We note that a time-dependent regulation of the proliferation rate $\pobp(t)$ is essential to allow \obp\ cells to (i) rapidly proliferate in early stages of osteoblastogenesis (when the density of pre-osteoblasts is low) and (ii) reach a controlled steady state. Mathematically, a rapid, exponential-like increase in the \obp\ population may occur from Eq.~\eqref{obp} whenever $\pobp(t)\!-\!\dobp(t)$ is positive and does not decrease too fast in time.\footnote{The density of \obp s at time $t$ has a contribution proportional to $\exp \!\big\{\int_0^t \d t'\, (\pobp\!-\!\dobp)(t')\big\}$. This contribution increases faster than any power law in time (exponential-like increase) provided that $(\pobp\!-\!\dobp)(t)$ is positive and does not decrease faster than or as fast as $\Order(1/t)$.} On the other hand,  one sees from Eq.~\eqref{obp} that a necessary condition for the \obp\ population to stay bounded and to converge to a meaningful steady-state (with finite, positive cell densities) is that
\begin{align}\label{PleqD}
    \mathcal{P}_\obp(t) - \mathcal{D}_\obp(t) < 0, \qquad t\to\infty.
\end{align}
The regulation of proliferation given in Eq.~\eqref{pobp} enables us to fulfill both requirements (i) and (ii).

\paragraph{Pre-osteoblasts generation: differentiation vs proliferation.}
Differentiation from \msc s and proliferation of pre-osteoblasts are two different biological mechanisms that enable the population of osteoblasts to reach the size required in a \bmu\ for functional remodelling. The relative proportion of these two mechanisms \textit{in vivo} has not been quantified experimentally. Proliferation is a mechanism that exponentially inflates any deviation in the original population size. Proliferation thus provides a sensitive control of the population and the potential for a quick response. Of course if this is the dominant mechanism for increasing the size of the \obp\ cell population, a small change in proliferation rate may lead to a very large change in the \obp\ cell population. We observe here that the more proliferation becomes dominant, the more difficult it becomes for the final cell population to be well-controlled, as a small change in the rate of proliferation leads to a large change in cell population. By contrast, differentiation of \msc s is a mechanism that influences the initial population of pre-osteoblasts. This provides a more stable mechanism for controlling \obp\ cell population, but this has the potential disadvantage of requiring the recruitment and maintenance of large numbers of \msc s. Clearly, if differentiation is large, then proliferation needs to be limited to reach the same population size.

For these reasons, it is helpful in the model to introduce the relative proportion of \obu\ differentiation and \obp\ proliferation as a parameter. We introduce the fraction $\nu$ such that the generation of \obp s in the steady state is achieved with a fraction $\nu$ by \obp\ proliferation and with a fraction $1\!-\!\nu$ by \obu\ differentiation. Denoting steady-state values by an overline, the total generation rate of \obp s in the steady state is given by $\sigmabar_\obp = \dobubar\,\obu + \pobpbar\,\obpbar$ (see Eq.~\eqref{obp}). The first term represents the contribution of \obu\ differentiation and should thus account for a fraction $1\!-\!\nu$ of $\sigmabar_\obp$. The second term represents the contribution of \obp\ proliferation and should thus account for a fraction $\nu$ of $\sigmabar_\obp$. To determine the values of $P_\obp$ and $D_\obu$ that satisfy this, we impose
\begin{align}
    \pobpbar \obpbar = \nu\ \sigmabar_\obp, \qquad \dobubar \obu = (1\!-\!\nu)\ \sigmabar_\obp, \label{prolif-differ-fractions}
\end{align}
and use the fact that $\sigmabar_\obp = \dobpbar \obpbar$ in the steady state. With Eqs.~\eqref{dobu},\eqref{dobp},\eqref{pobp}, one then has from Eq.~\eqref{prolif-differ-fractions}:
\begin{align}
    &P_\obp(\nu,\obpsat) = \nu\ D_\obp\ \pirep\Big(\tfrac{\ \overline{\tgfb}\ }{k^\tgfb_\obp} \Big) \Big( 1-\tfrac{\obpbar}{\obpsat} \Big)^{-1},\label{pobp-nu-obpsat}
    \\&D_\obu(\nu,\obpsat) = (1\!-\!\nu)\ D_\obp\ \frac{\pirep\Big( \tfrac{\ \overline{\tgfb}\ }{k^\tgfb_\obp} \Big)}{\piact\Big(\tfrac{\ \overline\tgfb\ }{k^\tgfb_\obu}\Big)}\ \frac{\,\obpbar\,}{\ \overline\obu\ }.\label{dobu-nu-obpsat}
\end{align}
Therefore, provided that $P_\obp = P_\obp(\nu,\obpsat)$ and $D_\obu = D_\obu(\nu, \obpsat)$ in Eqs.~\eqref{pobp} and~\eqref{dobu}, the system reaches for any value of $\nu$ a steady state characterised by the same cell densities $\obpbar, \obabar$ and $\ocabar$ and regulatory factor concentrations \tgfbbar, \ranklbar, \etc.\ as in Ref.~\cite{pivonka-etal-1} (despite the additional proliferation term in Eq.~\eqref{obp}).\footnote{For $\nu=0$, the model of Ref.~\cite{pivonka-etal-1} is retrieved, except for a correction in the production rate of \rankl, see Appendix~\ref{appx:model}.}

The parameter $\nu$ enables us to investigate how the relative occurrence of \obp\ proliferation vs \obu\ differentiation in osteoblastogenesis affects bone remodelling with a model calibrated against the same healthy-state properties. To understand how a dysregulation of \obp\ proliferation affects bone remodelling in an anabolic disease, we will set in Section~\ref{sec:properties}
\begin{align}
    &P_\obp = P_\obp(\nu,\obpsat) + \Delta P_\obp, \label{prolif-offset}
    \\&D_\obu = D_\obu(\nu,\obpsat), \label{differ-normal}
\end{align}
and study the effects of $\nu, \obpsat$ and of the proliferation rate `offset' $\Delta P_\obp$ (which accounts for dysregulation) on the steady state of the system.

\section{Properties of the model}\label{sec:properties}
The steady-state cell densities represented by the model correspond to physiological cell densities (averaged at the tissue level) of a normal, healthy adult whose skeleton undergoes remodelling. While a baseline of mesenchymal stem cells and hematopoetic stem cells is implicitly assumed, a bone remodelling event is not necessarily induced. Indeed, the system of ODEs~\eqref{oca}--\eqref{oba} governing the evolution of $\obp(t)$, $\oba(t)$, and $\oca(t)$ always admits vanishing bone cell densities as a solution, whatever the density of \obu s and of \ocp s.

Specific signalling is required to commit these stem cells to the osteoblastic and osteoclastic lineage. The induction of a bone remodelling event appears to be a complicated and poorly-understood process, that first requires bone lining cells retracting from the bone surface, and is followed by the recruitment of osteoclasts on site. Our model is not capable of modelling this induction process.  However, the specific signalling between osteoblasts and osteoclasts mediated by \rankl\ and \tgfb\ forms a positive feedback loop that leads any initial population of pre-osteoblasts or active osteoclasts to a steady state characterised by positive cell densitites $\obpbar, \obabar, \ocabar$ \cite{pivonka-etal-1,buenzli-pivonka-smith}.

\begin{figure}
    \centering\includegraphics[width=\figurewidth]{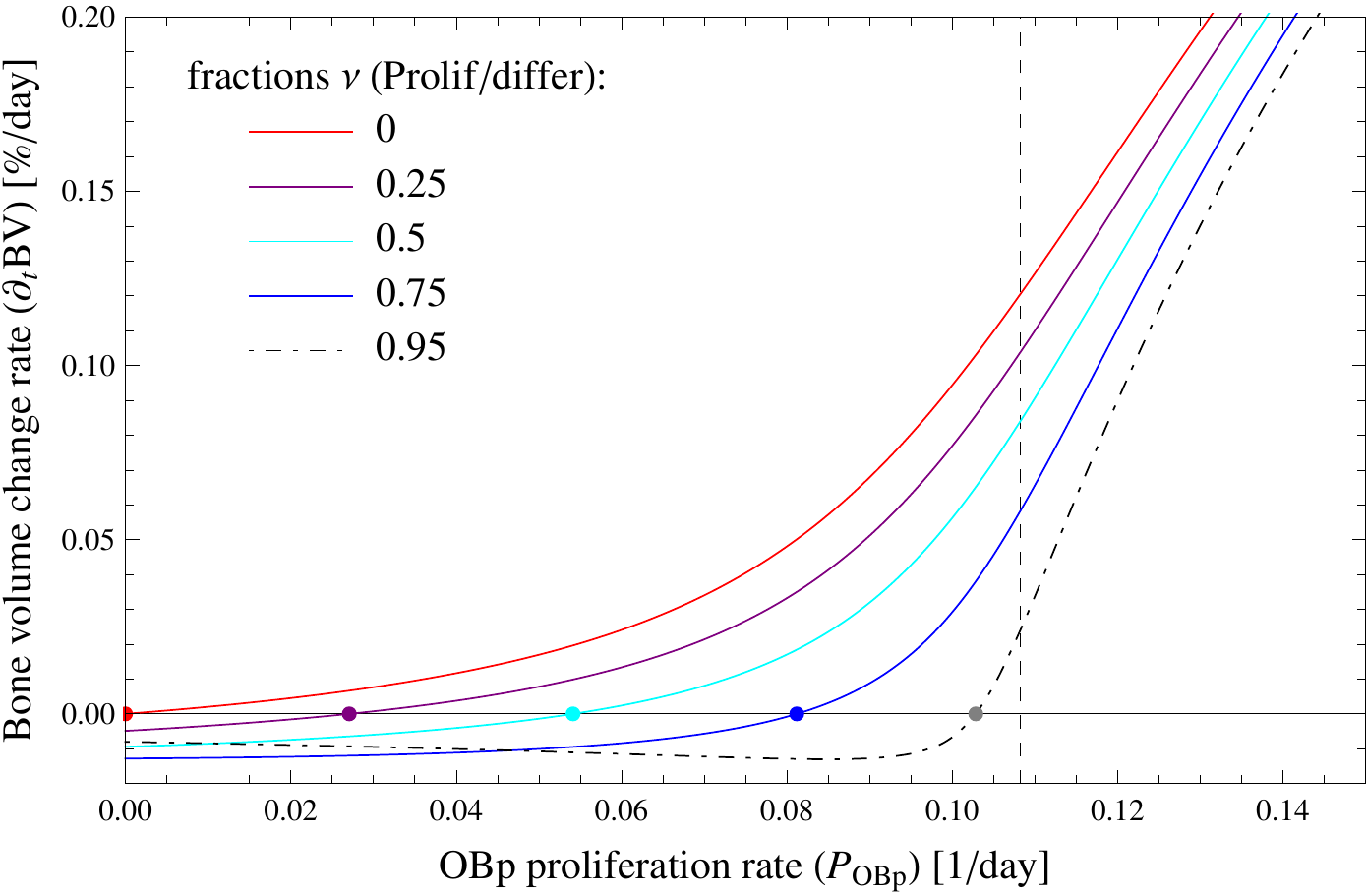}
    \caption{Steady-state value of bone volume change rate (in percent volume fraction/day) plotted against $P_\obp = P_\obp(\nu, \obpsat) + \Delta P_\obp$ for different fractions $\nu$ and a common value of $\obpsat = 0.04~\pM$. Each curve's zero is marked by a dot and represent the value $P_\obp(\nu,\obpsat)$ at which bone volume is steady (see Eq.~\eqref{prolif-offset}). The vertical dashed line represents the upper bound $P_\obp(1, \obpsat)$. }
        \label{fig:BVChangeRateVsOBpProlif}
\end{figure}

\paragraph{Anabolic potential of pre-osteoblast proliferation.}
Our previous investigations of the bone remodelling model of Ref.~\cite{pivonka-etal-1} have revealed that the \rank--\rankl--\opg\ pathway is effective at inducing catabolic behaviour in response to an increase in the \rankl/\opg\ ratio, but not effective at inducing anabolic behaviour in response to a decrease in the \rankl/\opg\ ratio \cite{pivonka-etal-2}. By contrast, Figure~\ref{fig:BVChangeRateVsOBpProlif} shows that increasing $P_\obp$ from a normal state with steady bone volume is very effective at inducing an anabolic behaviour of bone remodelling. But decreasing $P_\obp$ from this state is not effective at inducing a catabolic behaviour of bone remodelling, even at high fractions~$\nu$.

The strong anabolic potential of pre-osteoblast proliferation occurs despite pre-osteoblasts expressing \rankl, which by binding to the \rank\ receptor of osteoclasts promotes osteoclast activation. This is similar to the bone formative therapeutic strategy investigated by Lemaire~\etal~\cite{lemaire-etal}. In fact, the dynamics shows that active osteoclasts are only transiently increased by an increase in pre-osteoblast density. The increase in \obp s (which promotes osteoclastogenesis by increasing \rankl\ signalling to \ocp s) is followed by a delayed increase in \oba s. The latter cells produce \opg, which binds competitively to \rankl. This reduces the initial increase in \rankl\ signalling back to near-normal levels. Another limiting factor for osteoclastogenesis by \rankl\ signalling is the limited number of \rank\ receptors on \ocp s. The generation rate of active osteoclasts saturates when all \rank\ receptors on \ocp s are bound to \rankl.

It is noteworthy that the \rank--\rankl--\opg\ signalling patway exhibits a pronounced `catabolic bias' in the bone remodelling models developed by Lemaire \etal~\cite{lemaire-etal} and by Pivonka \etal~\cite{pivonka-etal-1,pivonka-etal-2}, while pre-osteoblast proliferation exhibit a ``complementary''  `anabolic bias' in the present model.  We emphasise that depending on the individual, such biases may not be as pronounced in practice as the models suggest. In the models, these biases can be partially explained by the rapid saturation of the receptor--ligand binding reaction rates (similarly to Michaelis--Menten enzyme kinetics) that limit the cells' response to extracellular ligands (via the `activator' and `repressor' functions $\piact$ and $\pirep$). As a consequence, cell behaviour is asymmetrical in response to an increase or to a decrease of extracellular ligands. The strength of this asymmetry depends on where on the curves $\piact$ and $\pirep$ the normal state is assumed to be. Normal ligand concentrations are likely to differ across indivuals. In some individuals, this normal ligand concentration may lie closer to the initial linear part, or final saturated part of the functions $\piact$, $\pirep$ than in other individuals, and in this way, lead to a less pronounced asymmetry of the cell's response.

\paragraph{Response to `continuous' \pth\ administration.}
The inclusion of \obp\ proliferation into the model introduces an additional mechanism for osteoblastogenesis. The relative importance of this additional mechanism is represented by the parameter $\nu$ introduced in Section~\ref{sec:model}. High fractions $\nu$ emphasise proliferation, which makes the model sensitive to small variations in the initial populations. Depending on the value of $\nu$, different system behaviours may arise, as illustrated in the following.

\begin{figure*}[h!]
        \centering\begin{tabular}{ll}
        \hspace{1mm}{\small (a)}{\small $\quad \nu=0.75, \quad\obpsat=0.005~\pM$}&\hspace{6mm}{\small (b)}{\small $\quad\nu=0.75, \quad\obpsat=0.01~\pM$}
        \\
        \hspace{-3mm}\includegraphics[width=0.49\textwidth]{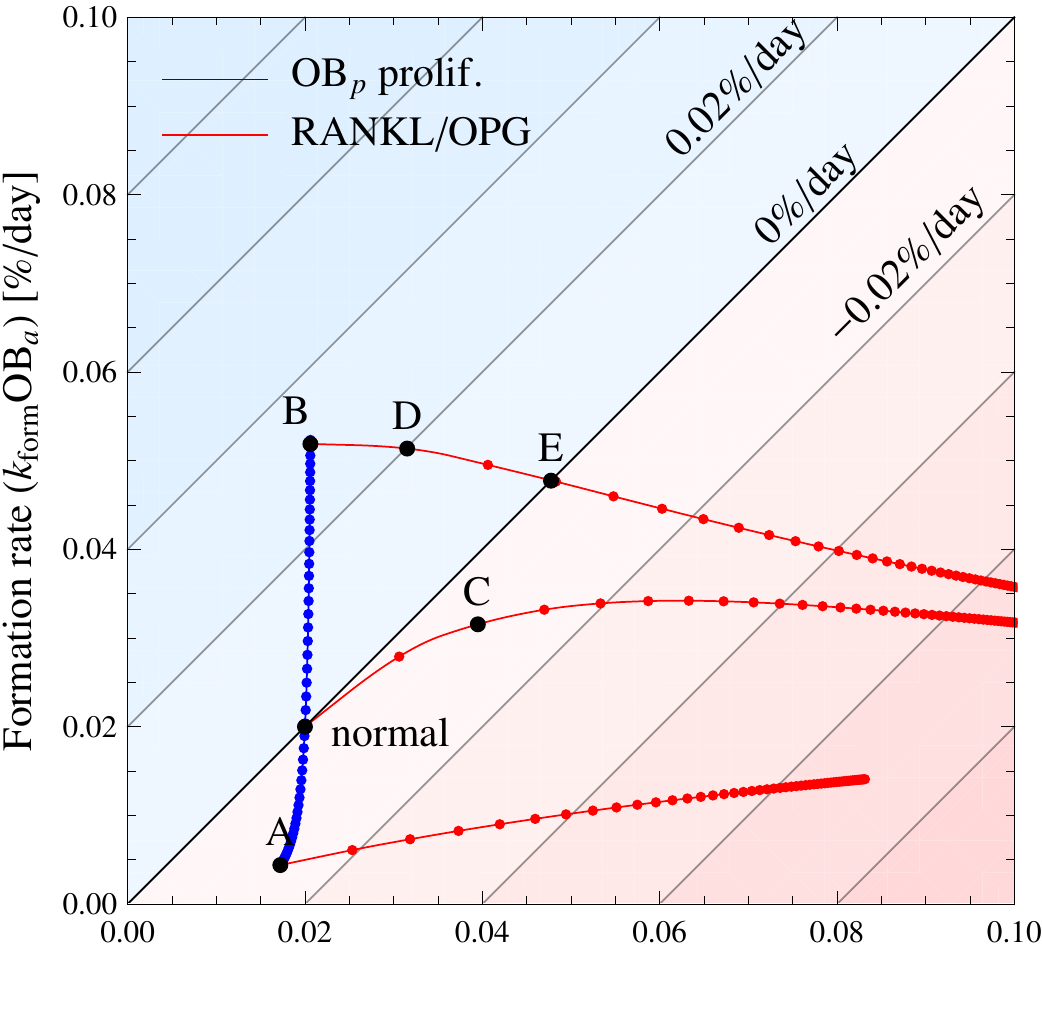}
        &
        \hspace{0mm}\includegraphics[width=0.49\textwidth]{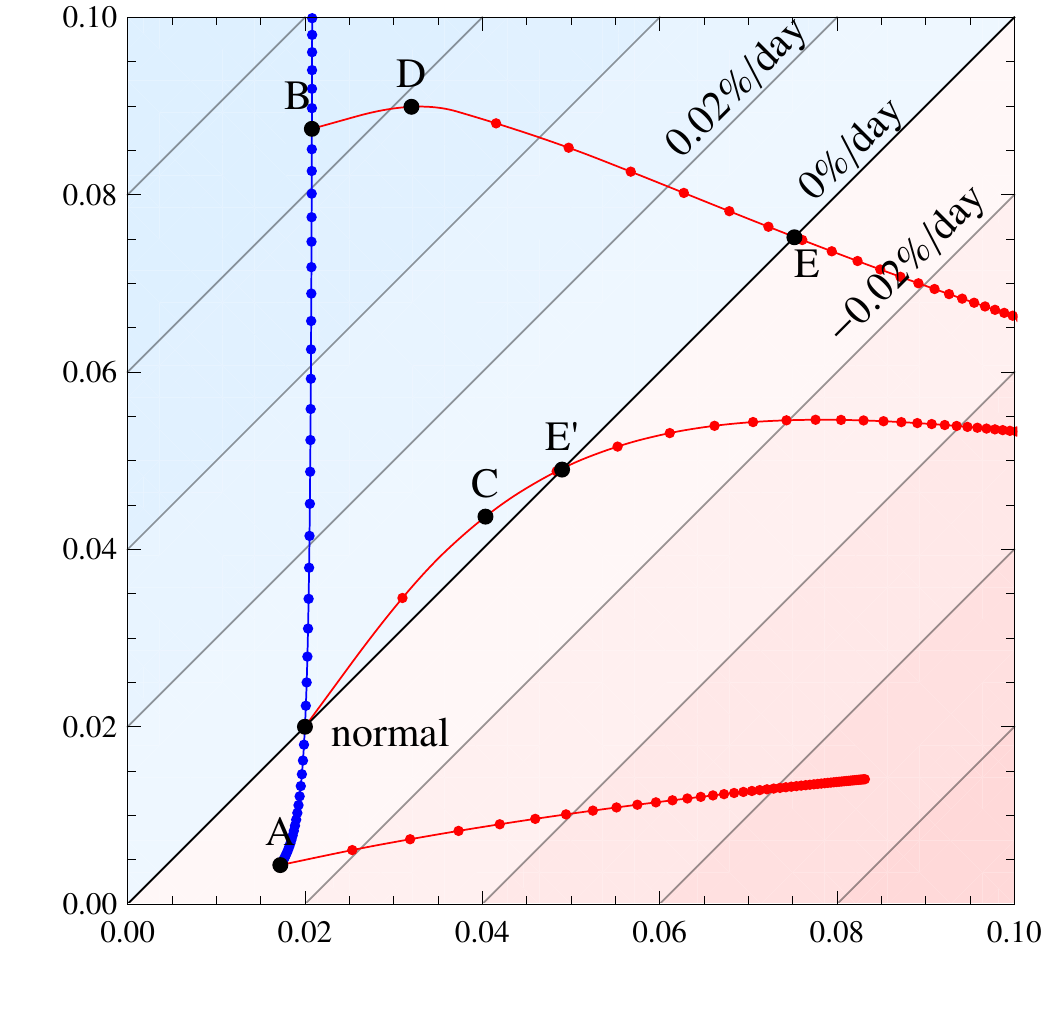}
        \\
        \hspace{1mm}{\small (c)}{\small $\quad\nu=0.75, \quad\obpsat=0.03~\pM$} &\hspace{6mm}{\small (d)}{\small $\quad\nu=0.25, \quad\obpsat=0.03~\pM$}
        \\
        \hspace{-3mm}\includegraphics[width=0.49\textwidth]{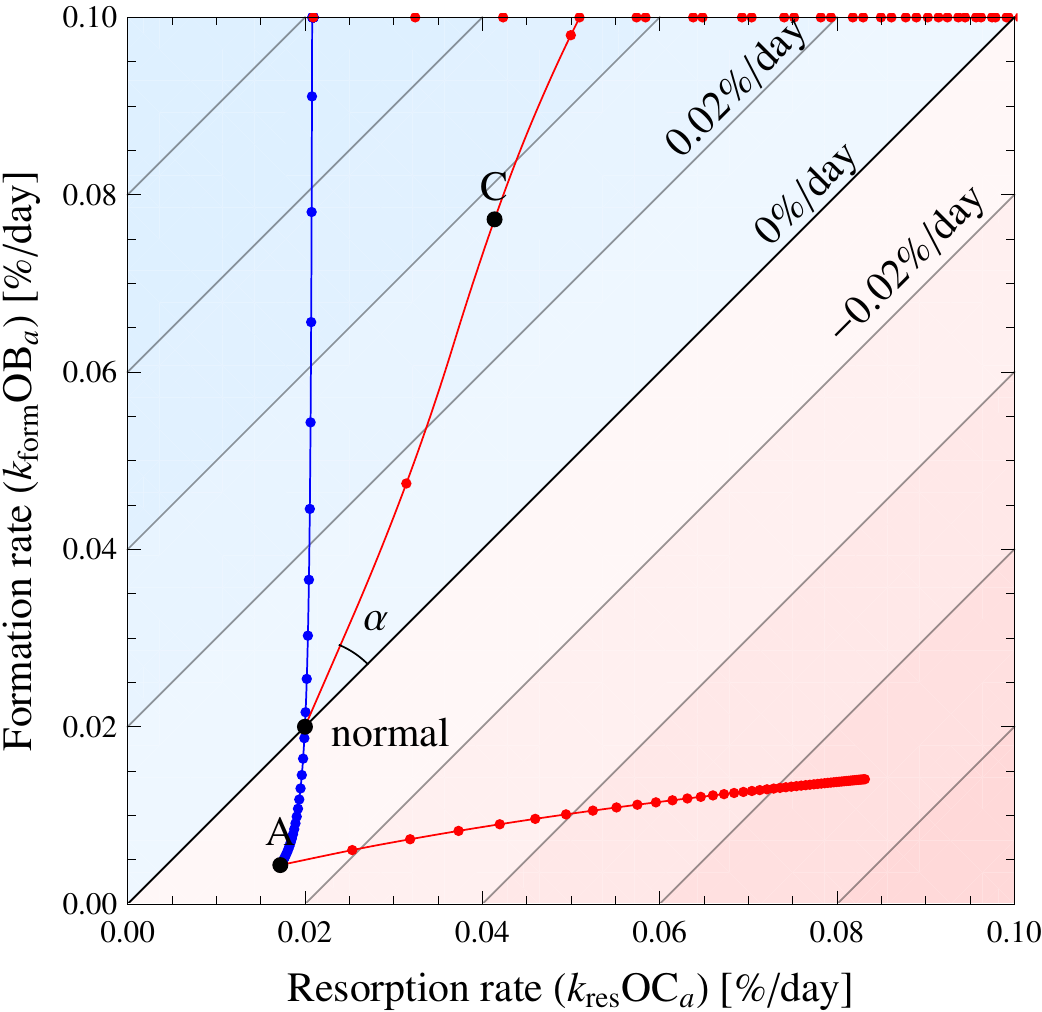}
        &
        \hspace{0mm}\includegraphics[width=0.49\textwidth]{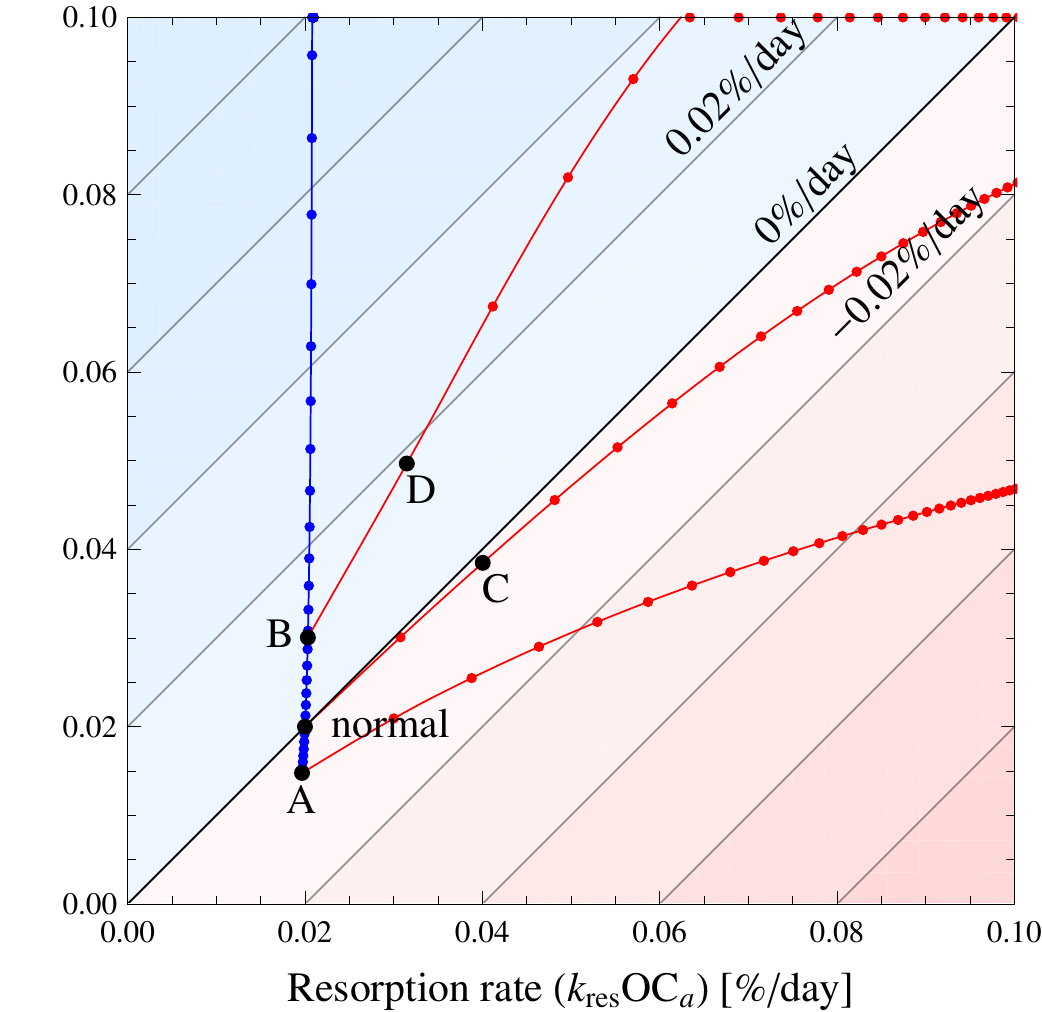}
    \end{tabular}
    \caption{Formation rate \vs\ resorption rate in the steady states obtained by varying \obp\ proliferation rates $P_\obp$ (blue curve) and by varying \rankl/\opg\ ratios (via continuous \pth\ administrations $\prodextpth$) (red curves). The various labelled points correspond to the following pairs $(P_\obp, \prodextpth)$: normal $\equiv \big(P_\obp(\nu,\obpsat), 0\big)$;  A $\equiv(0, 0)$; B $\equiv \big(2P_\obp(\nu,\obpsat), 0\big)$; C $\equiv\big(P_\obp(\nu,\obpsat),400/\da\big)$; and D $\equiv \big(2P_\obp(\nu,\obpsat),200/\da\big)$; Points E in (a), (b) and E' in (b) correspond to states with no bone gain nor loss, but higher turnover rate.}
\label{fig:OCaOBaPlot}
\end{figure*}

While an increase in pre-osteoblast proliferation is observed to induce a strong anabolic response for a broad range of fractions $\nu$ (Figure~\ref{fig:BVChangeRateVsOBpProlif}), the magnitude of the catabolic response to `continuous' \pth\ administration (which increases the \rankl/\opg\ ratio), is strongly dependent on the choice of $\nu$ and $\obpsat$.  Figure~\ref{fig:OCaOBaPlot} shows the steady-state resorption and formation rates reached by the model for four combinations of $(\nu, \obpsat)$ (Fig.~\ref{fig:OCaOBaPlot}(a)--(d)) under two externally-driven influences:
\begin{enumerate}
    \item[(i)] An altered value of the \obp\ proliferation rate parameter $P_\obp$ (blue curve).
    \item[(ii)] A continuous administration of \pth\ at rate $P^\text{ext}_\pth$  (see Eq.~\eqref{pth}) (red curves);
\end{enumerate}
In Figure~\ref{fig:OCaOBaPlot}, the alteration of \obp\ proliferation may represents an alteration in the \wnt\ pathway. Continuous \pth\ administration increases the concentration of \pth\ and thus increases the \rankl/\opg\ ratio (see Eqs.~\eqref{rankl}, \eqref{opg}), which promotes osteoclastogenesis. It is well-known that continuous \pth\ administration leads to a catabolic response. Strikingly, Figure~\ref{fig:OCaOBaPlot}(b) and (c) exhibit two examples of pairs $(\nu, \obpsat)$ that lead to an (as far as the authors are aware; unphysiological) anabolic behaviour in response to an increase in the \rankl/\opg\ ratio from the normal state. Decreasing $\obpsat$ and\slash or $\nu$ can restore the expected catabolic behaviour. This is seen by comparing Figure~\ref{fig:OCaOBaPlot}(b)$\to$(a) (decrease in \obpsat) and Figure~\ref{fig:OCaOBaPlot}(c)$\to$(d) (decrease in $\nu$).

\begin{figure}[t!]
    \centering\includegraphics[width=\figurewidth]{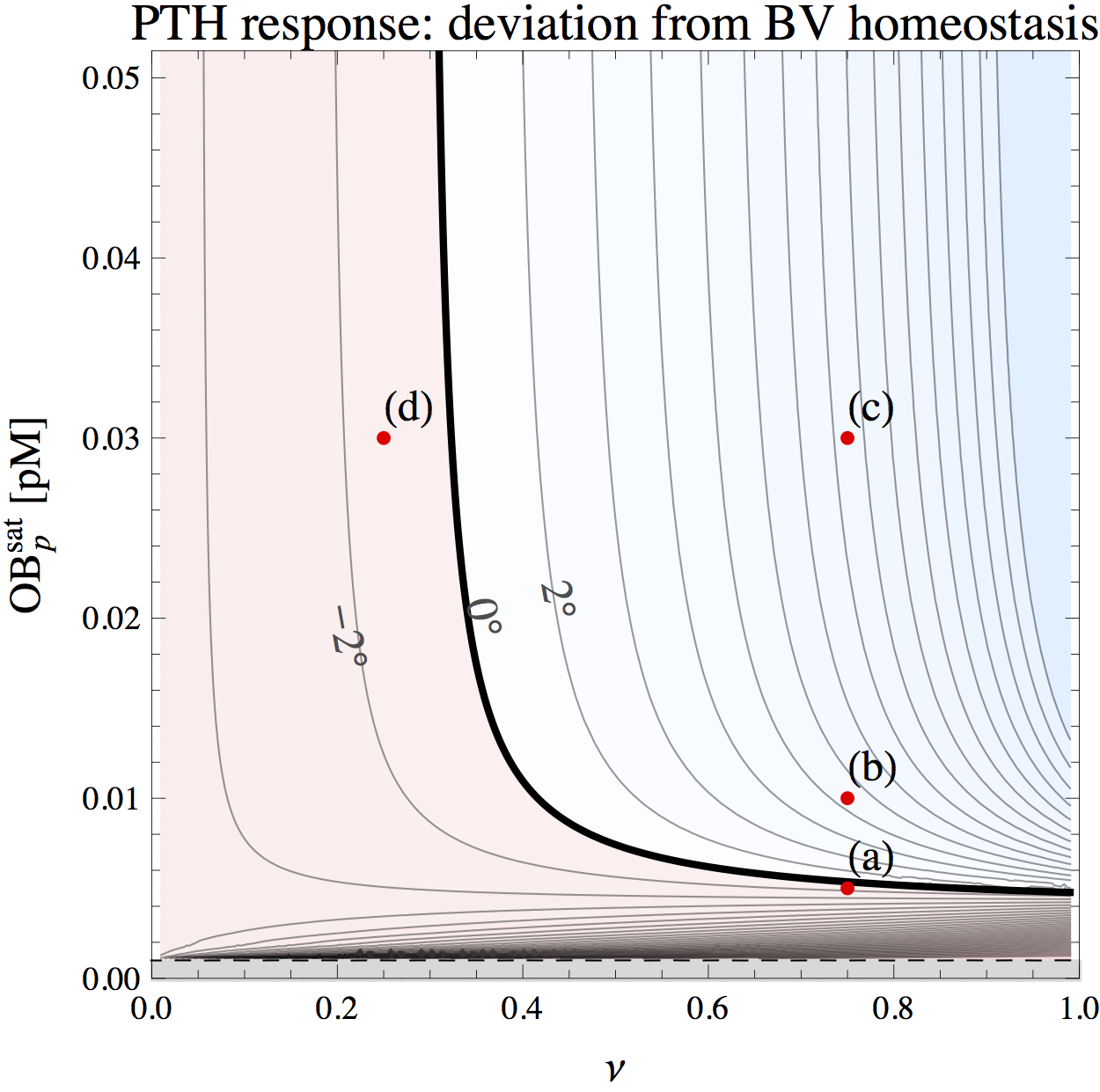}
    \caption{Angle between the continuous \pth\ administration response curve and the diagonal at the point corresponding to the normal state in Figure~\ref{fig:OCaOBaPlot}. Positive (negative) angles represent a response to increase in \pth\ directed towards anabolic (catabolic) states. The four situations (a)--(d) of Figures~\ref{fig:OCaOBaPlot} are also represented. Only negative angles (red region) represent the physiologically expected catabolic response to continuous \pth\ administration. The grey region is not part of the allowable parameter space as it corresponds to $\obpsat < \overline{\obp}$.}
    \label{fig:AnglePTHResponse}
\end{figure}
The possibility for such unphysiological anabolic behaviour is new compared to the models by Lemaire \etal~\cite{lemaire-etal} and Pivonka \etal~\cite{pivonka-etal-1}. However, this variability may be advantageous, allowing the system to be adjusted to specific patients or patient groups responding differently to increased \pth. The anabolic or catabolic behaviour of the model in response to increased \rankl/\opg\ ratio can be measured in Figure~\ref{fig:OCaOBaPlot} by the angle that the \pth\ curve makes with the diagonal (corresponding to \bv\ homeostasis) at the point corresponding to the normal state. This angle is shown in Figure~\ref{fig:OCaOBaPlot}(c) as `$\alpha$' and is plotted against $\nu$ and $\obpsat$ in Figure~\ref{fig:AnglePTHResponse}. Only the region corresponding to negative angles (bottom-left region, in red, in Figure~\ref{fig:AnglePTHResponse}) corresponds to a catabolic response to increased \pth. A physiologic estimate of this angle constrains $(\nu, \obpsat)$ to be on the contour line corresponding to this angle in Figure~\ref{fig:AnglePTHResponse}, leaving one degree of freedom. To retrieve the same catabolic behaviour to continuous \pth\ administration near the normal state for normal individuals as in Refs~\cite{pivonka-etal-1,pivonka-etal-2}, we choose this angle to be $\approx -4.5^\circ$. We note, however, that the catabolic response to continuous \pth\ in the present model is stronger at larger values of \pth\ administration rates (not shown).

While continuous \pth\ administration (infusion) does not induce an anabolic response, it is known that intermittent \pth\ administration (daily injections) does lead to an anabolic response. This dual catabolic--anabolic mode of action of \pth\ remains poorly understood~\cite{jilka}. It is instructive to understand within our model how an anabolic response to continuous \pth\ administration is obtained in Figure~\ref{fig:OCaOBaPlot}(b) and (c). This anabolic response of the model occurs when $\nu$ or \obpsat\ is large, \ie\ when \obp\ proliferation is significant. Increasing \pth\ increases \rankl/\opg\ and promotes osteoclastogenesis, which frees \tgfb\ in the microenvironment and increases the \obp\ population. If \obp\ proliferation is significant, this increase in \obp\ is amplified strongly and eventually overcomes \pth-induced osteoclastogenesis, which leads to an anabolic behaviour. As this behaviour is not observed \textit{in vivo} for continuous \pth\ administration, it can be expected that the proliferative potential of pre-osteoblasts is normally limited to the negative angle region in Figure~\ref{fig:AnglePTHResponse}. We estimate that the balance between \obu\ differentiation and \obp\ proliferation is probably somewhere in the range $0.4\lesssim \nu \lesssim 0.6$. This imposes a strong upper limit to the parameter \obpsat\ (see Figure~\ref{fig:AnglePTHResponse}). For an angle $\approx -4.5^\circ$, one has $\obpsat\lesssim 0.005~\pM$.

Finally, we note that intermittent \pth\ administration may exert an action on a variety of regulatory pathways of bone remodelling~\cite{jilka}. An overall anabolic response may be obtained as a combined effects of anabolic and catabolic disruptions of bone remodelling. This is the case for example of point D in Figure~\ref{fig:OCaOBaPlot}(a), where the superposition of an upregulation of pre-osteoblast proliferation and a catabolic response to \pth\ administration still leads to an overall anabolic response.

\section{Application to prostate cancer metastasis}\label{sec:pca}
Many bone pathologies are due to an altered bone balance and an altered bone turnover rate during remodelling. Bone imbalance is associated with under-refilling (bone loss) or over-refilling (bone gain) in \bmu s. Bone turnover rate is associated with the number of active \bmu s and indicates how fast bone may be lost, gained, and/or turned over. Our computational model represents bone remodelling at the tissue scale, where \bmu\ quantities are spatially averaged. At this scale, bone imbalance and abnormal turnover rates are characterised by altered overall rates of bone resorption $\kres\oca$ and bone formation $\kform\oba$ in the representative volume element~\cite{parfitt-in-recker}.

Prostate cancer develops metastases primarily to trabecular bone of the pelvis, femur and vertebral bodies~\cite{bubendorf-etal}. Several regulatory factors produced by the metastasising prostate cancer cells (\pca) interfere with the normal regulation of bone remodelling, leading to osteoblastic (anabolic) lesions with underlying osteolytic (catabolic) areas~\cite{keller-etal,clarke-fleisch,chirgwin-guise}. The molecules \wnt\ in particular, are believed to be particularly important in establishing osteoblastogenesis in these lesions~\cite{hall-etal-2005}. Hall \etal~\cite{hall-etal-2006} suggest that inhibition of \wnt\ by \dkk1\ at an early stage of \pca\ metastasis leads to osteolytic lesions (due to expression of \eg\ \pthrp\ or \rankl\ by the \pca\ cells). These lesions help the \pca\ cells to establish in the bone microenvironment. At a later stage, \pca\ cells progressively increase the \wnt/\dkk1\ ratio, resulting in an increased osteoblastic response. Prostate cancer cells also produce \psa, which cleaves \pthrp\ after amino acid 23~\cite{cramer-chen-peehl,keller-etal,chirgwin-guise,logothetis-lin}. The cleaved form \pthrpcleaved\ fails to activate the \pth\ receptor on osteoblasts, but is thought to promote osteoblastogenesis~\cite{chirgwin-guise}.

The above time course of metastatic bone lesions can be simulated in the model by prescribing an assumed time course for the population of \pca\ cells and for their expression of regulatory factors. To simplify, we assume that a \pca\ tumour implants itself in trabecular bone and locally grows over a characteristic time $\tau_\pca$ to a maximum density $\pcamax$:
\begin{align}
        \pca(t) = \pcamax\big[1-\exp(-t/\tau_\text{\pca})\big].\label{pca}
\end{align}
The \pca\ cells are assumed to produce \pthrp\ at a constant rate $\beta_\pthrp$, and \psa\ at a slowly increasing rate $\beta_\psa(t)$:
\begin{align}
    &\beta_\psa(t) = \beta_\psa^\text{max}\big[1-\exp(-t/\tau_\psa)\big],
\end{align}
The production rate of \wnt, $\beta_\wnt(t)$, is assumed low initially (or inhibited by \dkk1), but increases at later times:
\begin{align}
    \beta_\wnt(t) = \beta_\wnt^\text{max}  \frac{\e^{(t-t_\wnt)/\tau_\wnt} + \beta_\wnt^\text{min}/\beta_\wnt^\text{max}}{\e^{(t-t_\wnt)/\tau_\wnt} + 1}.\label{wnt-production}
\end{align}
In Eq.~\eqref{wnt-production}, $\beta_\wnt^\text{min}$ is the minimum production rate, $\beta_\wnt^\text{max}$ is the maximum production rate, $t_\wnt$ is the time at which the production rate increases and $\tau_\wnt$ measures the duration of that increase.

\begin{figure}[h!]
    \centering
    \includegraphics[width=\figurewidth]{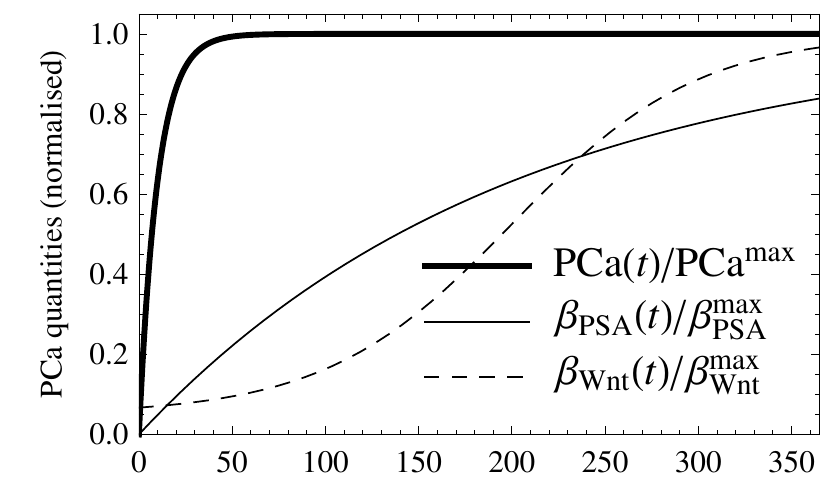}    
    \includegraphics[width=\figurewidth]{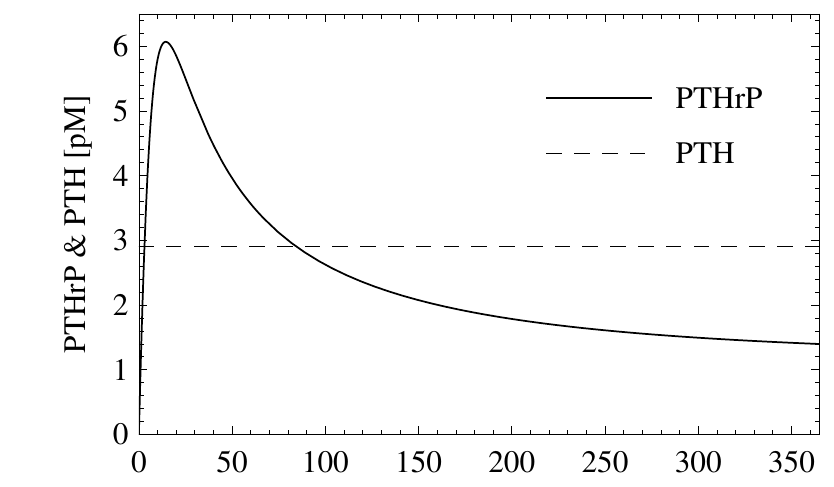}
    \includegraphics[width=\figurewidth]{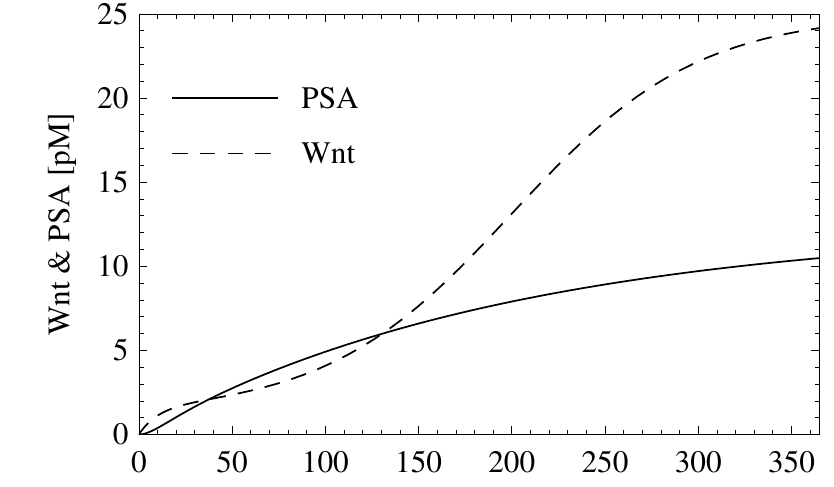}
    \includegraphics[width=\figurewidth]{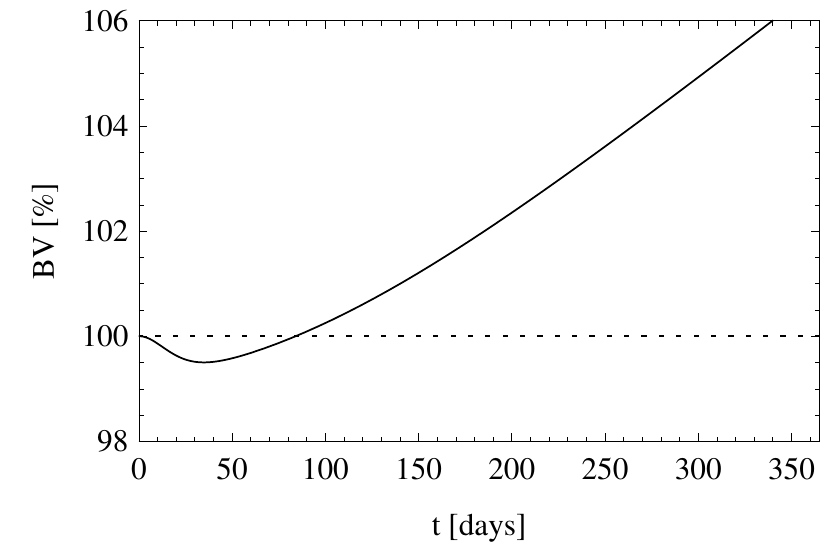}
    \caption{Time courses of prostate-cancer-induced $\pthrp(t)$, $\psa(t)$ and $\wnt(t)$, and their effect on the local bone volume fraction. The assumed evolution of the tumour and the rate of \psa\ and \wnt\ expression per \pca\ cell is seen in the topmost plot. The metastatic lesions transition from osteolytic to osteoblastic due to \wnt\ upregulating \obp\ proliferation.}
    \label{fig:metastatic-evolution}
\end{figure}
Both the catabolic influence of \pthrp\ and the anabolic influence of \wnt\ produced by the \pca\ cells are taken into account in the bone remodelling model. Cleavage of \pthrp\ by \psa\ is taken into account, but not a potential anabolic influence of \pthrpcleaved. Binding properties of $\pthrp$ on osteoblasts are assumed identical to those of w\pth, and so the concentration of $\pthrp$ is added to that of $\pth$ in Eqs.~\eqref{rankl},\eqref{opg}. The \wnt\ produced by the \pca\ cells is assumed to promote \obp\ proliferation according to:
\begin{align}
    P_\obp = P_\obp(\nu,\obpsat) \left[ 1 + \alpha^\wnt_\obp \piact\big(\wnt(t)/k^\wnt_\obp\big) \right]. \label{pobp-wnt}
\end{align}

As for the regulatory factors of bone remodelling, we assume that the binding reactions involving \pthrp, \psa\ and \wnt\ are fast. The concentration of these molecules quickly reaches a quasi-steady state equal to the production rate divided by the degradation rate~\cite[Eq.~(25)]{pivonka-etal-1}. We thus have:
\begin{align}
    &\pthrp(t) = \frac{\beta_\pthrp \pca(t)}{D_\pthrp + k^\psa_\pthrp \psa(t)}
    \\&\psa(t) = \frac{\beta_\psa(t) \pca(t)}{D_\psa}
    \\&\wnt(t) =\frac{\beta_\wnt(t) \pca(t)}{D_\wnt},\label{wnt}
\end{align}
where $D_\pthrp$, $D_\psa$ and $D_\wnt$ are degradation rates. Table~\ref{table:pca-parameters} in Appendix~\ref{appx:parameters} lists the parameter values associated to Eqs.~\eqref{pca}--\eqref{wnt}.

\begin{figure}[t]
    \centering
    \includegraphics[width=\figurewidth]{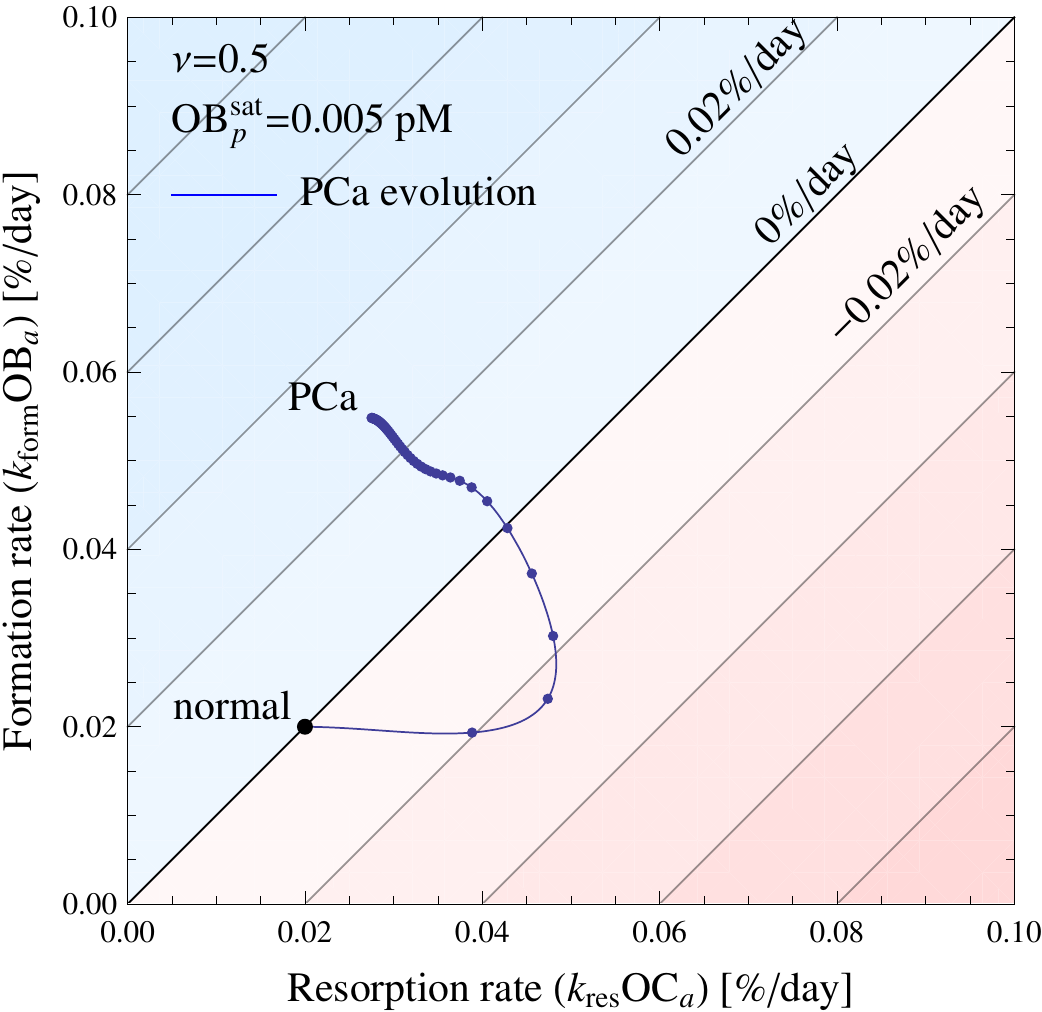}
    \caption{Simulated resorption rate and formation rate due to prostate cancer metastases to a tissue-scale portion of trabecular bone (blue line). The dots on the line mark the elapsed weeks. The metastatic lesion transitions from osteolytic to osteoblastic with increased turnover.}
    \label{fig:metastatic-resorption-formation}
\end{figure}

The time course of the concentrations $\pthrp(t)$ and $\wnt(t)$ in the bone microenvironment, and their effect on the local bone volume fraction, are shown in Figure~\ref{fig:metastatic-evolution}. It has to be emphasised that $\bv(t)$ does not represent the time course of the whole skeleton, but rather the evolution of a small part of trabecular bone within a tissue sample. Other regions of the bone might follow the same trend but have a different time course. This hypothesis is supported by histological evidence by Roudier \etal~\cite{roudier-etal} in patients who died with multiple bone metastases. This study shows that both regions of osteolytic lesions and regions of osteoblastic lesions are often found in the same individual. 

The joint evolution of the resorption rate and formation rate of this simulated disease are displayed in Figure~\ref{fig:metastatic-resorption-formation}. One sees that while the cancer develops osteoblastic lesions, turnover rate is higher than normal, and so resorption rate is also higher than normal. This is also consistent with the description by Clarke and Fleisch~\cite{clarke-fleisch} of prostate cancer lesions to bone being often a combination of both an increase in resorption and in formation at a same site. Here, we have driven the transition between osteolytic lesions and osteoblastic lesions by an increase in \wnt\ production around $t_\wnt$. While several other factors are known to influence the co-evolution of prostate cancer metastases and bone lesions, this sequence of events may already capture an aspect of metastatic lesions to bone, namely, that the interference of cancer-cell-produced cytokines with the normal biochemistry of bone remodelling can disrupt normal remodelling signals and drive it to either catabolic and anabolic imbalances.

\section{Conclusions}

Recent experimental evidence suggests that osteoblast proliferation plays an important role in the regulation of bone remodelling. In this paper, we have developed a novel computational model of bone cell interactions that includes osteoblast proliferation. This model takes into account a catabolic regulatory mechanism of bone remodelling, mediated by the \rank--\rankl--\opg\ pathway, and a new anabolic regulatory mechanism of bone remodelling, driven by osteoblast proliferation. From our numerical simulations the following observations have been made:
\begin{itemize}
    \item Preosteoblast proliferation has the potential for a strong anabolic bone response. Such a response could be mediated by a variety of signalling molecules including \wnt. The strong anabolic response of proliferation complements the strong catabolic response of \rankl\ observed in our model;
    \item To obtain physiologically meaningful results and a manageable control of osteoblastogenesis, a balance between osteoblast differentiation and proliferation is essential, as well as a feedback regulation of proliferation. This feedback regulation probably originates in the limited spatial and metabolic resources within the confines of the \bmu;
    \item Combining different strengths of pre-osteoblast proliferation with continuous \pth\ administration broadens the range of physiological bone responses that the model can represent. This may enable a better representation by the model of variability in the physiology of individuals.
    \item The example of prostate cancer metastasis to bone shows that the proposed catabolic and anabolic regulatory mechanisms of the model are able to simulate the progression of a complex bone disease ranging from catabolic to anabolic bone responses.
\end{itemize}

The numerical results indicate that the new model is improved and able to capture essential features of bone remodelling. Nevertheless, several aspects of the model can be further improved. In particular the phenomenological description of \wnt\ regulation of osteoblast proliferation could include biochemical binding reactions between different molecules regulating the binding properties of \wnt\ to its receptor \lrp{5/6}, such as sclerostin and \dkk1. Most interestingly, the variability of the bone response to a combination of continuous \pth\ administration and perturbation of osteoblast proliferation suggests that future developments of the model could shed light on the mechanisms underlying the difference between continuous \pth\ administration and intermittent \pth\ administration.

\begin{appendices}
\section{Rate equations of the regulatory factors}\label{appx:model}
The regulatory factor concentrations are governed by mass kinetics rate equations. Ligand--receptor binding reactions occur on a time scale much faster than the characteristic times of cellular response (such as differentiation, apoptosis). The rate equations for the regulatory factors can therefore be taken in their steady state (see Refs.~\cite{pivonka-etal-1,buenzli-pivonka-smith} for details). This gives:
\begin{align}
    &\tgfb(t) = \big[P^\text{ext}_\tgfb(t) + n_\tgfb^\text{bone} \kres \oca(t)\big]/D_\tgfb \label{tgfb}
    \\&\rankl(t) = \frac{P^\text{ext}_\rankl(t) + \beta^\rankl_\obp \obp(t)}{1 + k^\rankl_\rank \rank + k^\rankl_\opg \opg(t)}\notag
    \\&\phantom{\rankl(t)}\times\Bigg\{D_\rankl + \frac{\beta^\rankl_\obp \obp(t)}{N^\rankl_\obp \obp(t)\, \piact\big(\pth(t)/k^\pth_{\ob,\text{act}}\big)}\Bigg\}^{-1} \label{rankl}
    \\&\rank = N^\rank_\ocp\ \ocp, \label{rank}
    \\&\opg(t) = \frac{P^\text{ext}_\opg(t) + \beta^\opg_\oba\ \oba(t)\, \pirep\big(\pth(t)/k^\pth_{\ob,\text{rep}}\big)}{\beta^\opg_\oba\ \oba(t)\, \pirep\big(\pth(t)/k^\pth_{\ob,\text{rep}}\big)/\opg_\text{sat} + D_\opg} \label{opg}
    \\&\pth(t) = \big[P^\text{ext}_\pth(t) + \beta_\pth\big]/D_\pth \label{pth}
\end{align}
In these equations, external production rates $P^\text{ext}_L(t)$ represent external sources (or sinks) of the protein $L$ and are assumed given. We provide in Table~\ref{table:parameters} the description and values of the parameters of the model.

A slight change in the expression for \rankl\ in Eq.~\eqref{rankl} has been made compared to Ref.~\cite{pivonka-etal-1}. The production of \rankl\ is now correctly proportional to the number of cells that express \rankl. We have replaced $\beta_\rankl$ in Ref.~\cite[Eq. (36)]{pivonka-etal-1} by
$
    \beta^\rankl_\obp\,\obp(t).
$%
\footnote{We assume Model Structure 2 of Ref.~\cite{pivonka-etal-1}, in which \rankl\ is only expressed by \obp s and \opg\ is only expressed by \oba s.} To ensure that the normal steady state is unchanged by this correction, we take
$
    \beta^\rankl_\obp = \beta_\rankl/\obpbar. 
$
We note that the same inconsistency of having a production rate of \rankl\ not scaled by the number of osteoblasts is present in Ref.~\cite{lemaire-etal}. While many behaviours of the model are marginally affected by this correction, some inconsistent behaviours have been corrected. In particular, increasing the number of pre-osteoblasts in our model now increases the total number of \rankl\ (bound and unbound) accordingly, and transiently increases the number of active osteoclasts (until \opg, produced by \oba s, inhibits \rankl-activation of \rank). Previously, a decrease in the number of active osteoclasts was observed in this situation.

\section{Model parameters}\label{appx:parameters}
The parameters of the bone remodelling model are listed in Table~\ref{table:parameters}. The additional parameters introduced for the example of prostate cancer metastasis are listed in Table~\ref{table:pca-parameters}.
\begin{table*}
    \centering
        \vspace{-3mm}
    \caption{Model parameters}\label{table:parameters}
  \small
        \begin{tabular}{lrp{0.6\textwidth}@{}}
        \toprule
        Symbol & Value & Description
        \\\thickmidrule
        $\ocp$ & \num{1e-3}\,\pM & pre-osteoclast density
        \\$\ocabar$ & $\num{1e-4}\ \pM$ & steady-state density of active osteoclats
        \\$\obu$ & \num{1e-3}\ \pM & uncommitted osteoblast progenitors (\msc) density
        \\$\obpbar$ & $\num{1e-3}\ \pM$ & steady-state density of pre-osteoblasts
        \\$\obabar$ & $\num{5e-4}\ \pM$ & steady-state density of active osteoblats
        \\$n_\tgfb^\text{bone}$ & $\num{1e-2}\ \pM$ & density of \tgfb\ stored in the bone matrix
        \\$\kres$ & 200 $\pM^{-1}\da^{-1}$ & daily volume of bone matrix resorbed per osteoclast
        \\$\kform$ & 40 $\pM^{-1}\da^{-1}$ & daily volume of bone matrix formed per osteoblast
        \\\midrule
        $D_\ocp$ & $2.1/\da$& $\ocp\to\oca$ differentiation rate parameter
        \\$A_\oca$ & $5.65/\da$& \oca\ apoptosis rate parameter
        \\$D_\obu$ & $0.7/\da$& $\obu\to\obp$ differentiation rate parameter, value for $\nu=0.5$; Eq.~\eqref{differ-normal}
        \\$D_\obp$ & $0.166/\da$& $\obp\to\oba$ differentiation rate parameter
        \\$P_\obp$ & $0.054/\da$& \obp\ proliferation rate parameter, value for $\nu=0.5$; Eq.~\eqref{prolif-offset}
        \\$A_\oba$ & $0.211/\da$& \oba\ apoptosis rate
        \\\midrule
        $k^\rankl_\ocp$ &16.65 \pM& parameter for \rankl\ binding on \ocp
        \\$k^\tgfb_\oca$ &\num{5.63e-4}\ \pM& parameter for \tgfb\ binding on \oca
        \\$k^\tgfb_\obu$ &\num{5.63e-4}\ \pM& parameter for \tgfb\ binding on \obu
        \\$k^\tgfb_\obp$ &\num{1.75e-4}\ \pM& parameter for \tgfb\ binding on \obp
        \\$k^\pth_{\ob,\text{act}}$ &150\ \pM& parameter for \pth\ binding on \ob\ (for $\piact$)
        \\$k^\pth_{\ob,\text{rep}}$ &0.222\ \pM& parameter for \pth\ binding on \ob\ (for $\pirep$)
        \\$k^\rankl_\rank$ &0.034/\pM& association binding constant for \rankl\ and \rank
        \\$k^\rankl_\opg$ &0.001/\pM& association binding constant for \rankl\ and \opg
        \\$\beta^\rankl_\obp$ &$\num{1.68e5}/\da$& production rate of \rankl\ per \obp
        \\$\beta^\opg_\oba$ &$\num{1.63e8}/\da$& production rate of \opg\ per \oba
        \\$\beta_\pth$ &$250\,\pM/\da$& production rate of systemic \pth
        \\$N^\rankl_\obp$ & \num{2.7e6}& maximum number of \rankl\ per \obp
        \\$N^\rank_\ocp$ & \num{1e4}& number of \rank\ receptors per \ocp
        \\$\opg_\text{sat}$ & $\num{2e8}\,\pM$ & \opg\ density at which endogeneous production stops
        \\$n^\text{bone}_\tgfb$ & $0.01\,\pM$& density of \tgfb\ stored in the bone matrix
        \\$D_\tgfb$ &$2/\da$& degradation rate of \tgfb
        \\$D_\rankl$ &$10/\da$& degradation rate of \rankl
        \\$D_\opg$ &$0.35/\da$& degradation rate of \opg
        \\$D_\pth$ &$86/\da$& degradation rate of \pth
        \\\midrule
        $\nu$ & 0.5 & fraction of \obp\ proliferation over \obu\ differentiation involved in the steady-state density $\obpbar$
        \\$\obpsat$ & 0.005\ \pM & \obp\ density at which proliferation stops
        \\\bottomrule
    \end{tabular}
\end{table*}

\begin{table*}
    \centering
    \caption{\pca-specific parameters}\label{table:pca-parameters}
  \small
        \begin{tabular}{lrl}
        \toprule
        Symbol & Value & Description
        \\\thickmidrule
        \pcamax & \num{5e-3} \pM & maximum \pca\ density
        \\\midrule
        $\tau_\pca$ & 10 \days & duration of local \pca\ growth
        \\$\tau_\psa$ & 200 \days & duration of increase in \psa\ production
        \\$\tau_\wnt$ & 50 \days & duration of increase in \wnt\ production
        \\\midrule
        $t_\wnt$ & 200 \days & time of increase in \wnt\ production
        \\\midrule
        $\beta_\pthrp$ & \num{2e5}/\da& production rate of \pthrp\ per \pca
        \\$\beta_\psa^\text{max}$ & \num{1e4}/\da& final production rate of \psa\ per \pca
        \\$\beta_\wnt^\text{min}$ & \num{5e2}/\da& minimum production rate of \wnt\ per \pca
        \\$\beta_\wnt^\text{max}$ & \num{1e4}/\da& maximum production rate of \wnt\ per \pca
        \\\midrule
        $D_\pthrp$ &86/\da& degradation rate of \pthrp
        \\$D_\psa$ &4/\da& degradation rate of \psa
        \\$D_\wnt$ &2/\da& degradation rate of \wnt
        \\\midrule
        $k^\psa_\pthrp$ & $60\ \pM^{-1}\da^{-1}$ & parameter for \psa\ cleaving \pthrp
        \\$k^\wnt_\obp$ & $2\ \pM$ & parameter for \wnt\ binding on \obp
        \\\midrule
        $\alpha^\wnt_\obp$ & 2 & amplification factor of \pca-induced \obp\ proliferation
        \\\bottomrule
        \end{tabular}
\end{table*}
\clearpage
\end{appendices}

\end{document}